\documentclass[1p,preprint]{elsarticle}
\usepackage{physics,bm}
\usepackage{graphicx}

\usepackage{color}

\newcommand{\eref}[1]{(\ref{#1})}
\renewcommand{\d}{\dagger}
\newcommand{\HH}{\hat{H}}
\newcommand{\HQ}{\hat{Q}}
\newcommand{\Hp}{\mathrm{H}}
\newcommand{\Hc}{\mathcal{H}}
\newcommand{\Lc}{\mathcal{L}}
\newcommand{\Mc}{\mathcal{M}}
\newcommand{\Bc}{\mathcal{B}}

\newcommand{\Cc}{\mathcal{C}}
\newcommand{\Sb}{\Bar{S}}

\newcommand{\Xc}{\mathcal{X}}

\newcommand{\at}{\Tilde{a}}

\newcommand{\xit}{\Tilde{\xi}}

\newcommand{\ab}{\Bar{a}}

\newcommand{\xib}{\Bar{\xi}}

\newcommand{\nd}{\dot{n}}
\newcommand{\md}{\dot{m}}
\newcommand{\Nd}{\dot{N}}

\newcommand{\expint}[5]{e^{ {#1}\int^{#2}_{#3} \dd{#5} {#4} }}

\newcommand{\expvac}[1]{\mel{0}{#1}{0}}

\bmdefine{\bx}{x}
\bmdefine{\bk}{k}
\bmdefine{\by}{y}
\bmdefine{\bz}{z}
\bmdefine{\bq}{q}

\newcommand{\ve}{\varepsilon}

\newcommand{\intx}{\int\! d^3x\;}

\newcommand{\Sbi}[1]{\Sb^{+,#1}_{\mathrm{int},q}}

\newcommand{\DiagramA}{\scalebox{1}{%
		\begin{picture}(0.8,0.4)(0,-0.3) 
		\put(0, 0.2){\line(1,0){0.2}} %
		\put(0.6,0.2){\line(1,0){0.2}} %
		\put(0.4,0.2){\circle{0.4}} %
		\end{picture}}
}

\begin{document}

\begin{frontmatter}

\title{Nonequilibrium Thermo Field Dynamics Using $4\times 4$-Matrix Transformation 
for System with Bose--Einstein Condensation}

\author[densi]{T.~Oyama}
\ead{takahiro.o@aoni.waseda.jp}

\author[ICM,nagano]{Y.~Nakamura}
\ead{yusuke.n@asagi.waseda.jp}
\author[densi]{Y.~Yamanaka}
\ead{yamanaka@waseda.jp}

\address[densi]{Department of Electronic and Physical Systems, Waseda
University, Tokyo 169-8555, Japan}
\address[ICM]{Institute of Condensed-Matter Science, Waseda
University, Tokyo 169-8555, Japan} 
\address[nagano]{Nagano Prefectural Matsumoto Agatagaoka High School, 
Nagano 390-8543, Japan}

\begin{abstract}
This study aims to construct a theoretical formulation of a nonequilibrium process 
for a system of Bose--Einstein condensate associated with a spontaneous symmetry breakdown. 
For this, Thermo Field Dynamics is used. We then describe the most general forms of 
a $4\times 4$ transformation and an unperturbed Hamiltonian. 
After calculating the $4\times 4$ self-energy and identifying its on-shell, 
we impose the renormalization condition in which the entire on-shell self-energy should vanish. 
This condition provides a sufficient number of independent equations to determine all of the parameters 
in an unperturbed Hamiltonian, among which the quantum transport equations 
for normal and anomalous number distributions are included. 
\end{abstract}

\begin{keyword}
Thermal field theory \sep 
Thermo Field Dynamics \sep 
Nonequilibrium \sep 
Renormalization \sep 
Quantum transport equation \sep 
Cold atom

\end{keyword}

\end{frontmatter}

\section{Introduction}

A spontaneous symmetry breakdown (SSB) in quantum field theory is a mechanism 
that creates versatile and complex phenomena in the real macroscopic world 
from a simple microscopic theory or Hamiltonian with high symmetry. 
This is universal, from relativistic high-energy physics and cosmology 
to condensed matter physics, e.g., ferromagnetism, superfluidity, and 
superconductivity. Establishing a sound theoretical approach to nonequilibrium 
phenomena involving phases associated with an SSB is a challenging subject. 
In this connection, experiments on the Bose--Einstein condensate (BEC) of a cold atomic 
system under a trapping potential, which is a typical example of an SSB of a 
global phase symmetry \cite{PethickSmith,GriffinBook,VitielloBook}, 
are ideal for testing theoretical formulations and calculations of 
the nonequilibrium phenomena with an SSB \cite{Jin,Yamashita,Morgan1,Morgan2,Bezett},
because they are dilute and 
weakly interacting, and their nonequilibrium processes are extremely slow and 
can be observed directly \cite{Davis,Miesner,Sommer,LesHouches}.

Two formalisms of a nonequilibrium quantum field
 system are known, i.e., a closed time path or Keldysh--Schwinger
method \cite{Schwinger,Keldysh,KadanoffBaym,Danielewiecz,Chou}, 
and the other is Thermo Field Dynamics (TFD) \cite{UMT,AIP}. The approach 
in this paper is based on TFD,
in which a thermal average is given as an expectation of the 
thermal vacuum by doubling every degree of freedom. 
A $2 \times 2$-matrix thermal Bogoliubov transformation,
connecting physical operators to thermal quasiparticle operators that define the Fock space,
becomes time-dependent in nonequilibrium TFD. 
Starting from the choice of an appropriate unperturbed Hamiltonian with several parameters,
i.e., an energy shift and a time-derivative of the particle number distribution,
in the interaction picture and calculating the full propagator or self-energy in a 
$2 \times 2$ matrix form, we
determine the parameters from the self-consistent renormalization conditions.
As such, the quantum transport equation is derived from the renormalization condition.
When a BEC is created, we must consider the usual and thermal Bogoliubov 
transformations simultaneously in TFD, namely, TFD using a $4 \times 4$-matrix. 

Several studies on nonequilibrium TFD for BEC systems have been conducted. 
For a system with a stationary BEC and time-dependent number distribution, we 
formulated nonequilibrium TFD in Ref.~\cite{AnnPhys325} using a direct product of the two 
(usual and thermal) Bogoliubov transformations, 
each of which is represented by a $2 \times 2$ matrix. The quantum transport equation is
derived from the diagonalization condition in Ref.~\cite{ChuUmezawa}, meaning
 that the off-diagonal element of the full propagator, 
with respect to the thermal index of the doubling,
should vanish. In Ref.~\cite{AnnPhys326}, this study was extended to a nonequilibrium 
system with a time-dependent order parameter, given by the expectation of the Heisenberg 
field operator by means of a field expansion in the complete set of time-dependent 
wave functions. The quantum transport equation was obtained again from
the diagonalization condition, whereas the renormalization condition on the self-energy
fixed the energy shift.
 
Meanwhile, it was pointed out in Ref.~\cite{AnnPhys331} 
that the diagonalization condition on the full propagator, 
under its equilibrium limit, gives the same results as the equilibrium theory 
only in the lowest order loop-calculation, but never in higher orders. 
Instead of the diagonalization condition, we proposed a new renormalization condition
on the self-energy, which is a generalization of the on-shell renormalization
condition in quantum field theory at zero and finite temperatures and
yields a correct equilibrium limit. The energy shift of the quasiparticle and 
quantum transport equation follow from this self-energy renormalization in a unified manner.
The idea of self-energy renormalization has been extended in Ref.~\cite{IJMPKuwahara} 
to an inhomogeneous system such as the trapped cold atomic gas with a BEC,
for which the energy counter term is non-diagonal 
owing to a lack of the conservation of momentum.

In this study, we reconsider nonequilibrium TFD for a system with a
BEC from the following three points. As the first point, the most general 
$4\times 4$ matrix form of the unperturbed Hamiltonian 
having as many parameters as possible should be asked. In previous studies
\cite{AnnPhys325,AnnPhys326,IJMPKuwahara}, the forms of the unperturbed Hamiltonian 
have been restricted to a certain extent from the beginning, and 
the degree of freedom of the parameters in a $4\times 4$-matrix 
has not been fully exhausted. We note that 
 a possible imaginary part of the energy parameter was suppressed there, although
it naturally appears as a thermal instability under a thermal situation. 
At an early stage of developing nonequilibrium TFD, its general $4\times 4$-matrix
structure has already been explored in Ref.~\cite{Elmfors}. As the argument in 
Ref.~\cite{Elmfors} covered
only matrices with real elements, it is not related directly
to our present study, where imaginary parts of 
the parameters in a $4\times 4$-matrix are important. The second is concerned with treatments of 
the time dependences of the parameters, whereas the time derivatives of the number distributions are 
present as the counter terms in the unperturbed Hamiltonian, and we introduce
the time-dependent complete set (TDCS) to expand the field operator \cite{Matsumoto2001}, 
such that the quasiparticle 
operators do not depend on time.
The third point
is a question regarding how an on-shell part of the self-energy of a $4\times 4$-matrix is identified, 
particularly because the renormalized energy parameter is time-dependent and, 
moreover, complex. The presence of the imaginary part of the energy parameter
was not explicitly considered in the previous definitions. We can see
that an on-shell part can be defined irrespective of the presence of an imaginary part
owing to the introduction of a TDCS. Thus, the renormalization condition in which
the entire on-shell self-energy should vanish provides a sufficient number of independent equations
to determine all parameters in the unperturbed Hamiltonian.

To accomplish the objective above, in this study, we consider a homogeneous system 
of a quantum bosonic field with a BEC, thereby avoiding subtlety 
of the zero-mode, using a Bogoliubov approximation, 
and having no off-diagonal counter terms \cite{IJMPKuwahara}. 
For an application of nonequilibrium TFD to the trapped cold atomic
systems during the experiments, we must consider the discrete zero-mode \cite{NTY} and 
off-diagonal components of the counter terms.  
Furthermore, it is assumed that the order parameter is time-independent.
Although a time-dependent order parameter is more interesting, we
study the time-independent case as a first step here.

The remainder of this paper is organized as follows. In Sec.~\ref{sec-RevTFD},
a TFD formulation is briefly reviewed for later discussions.  In Sec.~\ref{sec-44Formulation}, we 
derive the most general forms of a $4\times 4$ transformation and the unperturbed 
Hamiltonian of TFD for a single mode. A method
applying a  TDCS is also presented.
Next, a model of quantum field theory is introduced in Sec.~\ref{sec-ModelQFT},
and the formulation described in Sec.~\ref{sec-44Formulation} transferred to this
 model is then described in Sec.~\ref{sec-NETFDQFT}. Section~\ref{sec-FeynmanSE} covers
 the Feynman diagram method in nonequilibrium TFD and calculations of the self-energies.
In Sec.~\ref{Sec:renormalization}, we describe the
 on-shell parts of the self-energies and the renormalization conditions imposed
on them to derive the quantum transport equations and
determine all parameters involved in the unperturbed Hamiltonian. Finally, Sec.~\ref{sec-Conclusion}
provides some concluding remarks and a summary of this study.

\section{Review of conventional Thermo Field Dynamics formulation}\label{sec-RevTFD}

In this section, we briefly review the formulation of TFD \cite{AIP, AnnPhys331},
using a bosonic system of a single mode,
i.e., $a$-operator, with 
an equal-time canonical commutation relation $\qty[ a, a^\d] = 1$. In TFD, 
thermal expectation $\mathrm{Tr}\qty[ \rho A]$ 
is replaced with the pure state expectation $\ev{A}{0}$\,, where
$A$ and $\rho$ represent an arbitrary function of the $a$-operator, and a normalized 
density matrix, respectively. The states $\bra0$ and $\ket0$, called 
the thermal vacua, are annihilated 
by the thermal quasiparticle operators, $\xi$ and $\xit$, 
\begin{align} \label{vacua-xi}
	\xi\ket0= \xit\ket0 =0 \,,\qquad
	\bra0\xi^\d = \bra0\xit^\d =0 \,.
\end{align}
Our main interest is to obtain a transport equation
that describes the time-evolution of the nonequilibrium number 
distribution $n(t)$, which is defined by
\begin{align}\label{numberdistribution}
	n(t) = \mathrm{Tr}\qty[  a^\d a \rho] = \ev{a^\d a}{0} \,.
\end{align}
The $a$- and $\xi$-operators are related through  
the thermal Bogoliubov transformation, which is generally not unique. For nonequilibrium TFD, we ordinarily take the 
$\alpha = 1$ representation \cite{AIP} because the critical concept 
of the thermal causality is naturally established \cite{AnnPhys331,IJMPKuwahara}.
The Bogoliubov transformation in the 
$\alpha=1$ representation is
\begin{align}\label{thermalBogoiubov1}
	\pmqty{a \\ \at^\d} = \pmqty{1 & n \\ 1 & 1+n} \pmqty{\xi \\ \xit^\d} \,,
\end{align}
under the assumption that the global phase symmetry is not broken. These operators satisfy 
equal-time commutation relations for non-tilde and tilde operators, respectively,
\begin{align}\label{ccr}
	&\qty[ a, a^\d] = \qty[ \at, \at^\d] = 1\,,\quad
	\qty[ a, \at] = \qty[ a, \at^\d] = 0 \,,\\
	&\qty[ \xi, \xi^\d] = \qty[ \xit, \xit^\d] = 1\,,\quad
	\qty[ \xi, \xit] = \qty[ \xi, \xit^\d] = 0 \label{ccr2}\,.
\end{align}
The feature of the $\alpha = 1$ representation is seen in the following relations of $\bra{0}$:
\begin{align}\label{TSC-a}
	\bra0 a = \bra0\at^\d \,,\qquad
	\bra0 \at = \bra0 a^\d \,.
\end{align}
Note that $\ket{0}$ is not a Hermitian conjugate of $\bra{0}$. 
In addition, $\xi^\d$ is not a Hermitian conjugate of $\xi$,
because the thermal Bogoliubov transformation is non-unitary.
The doubling of the operators in TFD is well explained in terms of the superoperator formalism 
\cite{SchmutzSO}, as has been discussed in Ref.~\cite{AnnPhys331}. Non-tilde and tilde operators 
correspond to the operations applied to the density matrix from left and right, respectively. 
They are related to each other according to the following tilde conjugation rules,
\begin{align} \label{tilde-rule1}
	& \qty(a)^\sim = \at \,,\qquad
	 \qty(\at)^\sim = a  \,,\qquad 
	 \qty(A^\d)^\sim = \qty(\tilde{A})^\d \,,\\
	&\qty( A_1 A_2)^\sim = \tilde{A_1}\,\tilde{A_2}\,, \qquad
	\qty(c_1 A_1 + c_2A_2)^\sim = c_1^* \tilde{A}_1 + c_2^* \tilde{A}_2 \,,\label{tilde-rule2}\\
	&\qty(\bra0)^\sim = \bra0 \,,\qquad \qty(\ket0)^\sim = \ket0 \,.\label{tilde-rule3}
\end{align}
The Heisenberg equations for 
the doubled operators are
\begin{align}\label{HeisenbergEq}
	i\frac{d}{dt} a_\Hp = \qty[ a_\Hp\,, \HH] \,,\qquad
	i\frac{d}{dt} \at_\Hp = \qty[ \at_\Hp\,, \HH] \,,
\end{align}
where $\HH$, called a hat Hamiltonian, is defined as $\HH = H - \tilde{H}$\,, 
which comes from the Liouville--von Neumann equation $i\pdv{t} \rho = H \rho - \rho H$\,.
Throughout this paper, $\hbar$ is set to unity.

The total Hamiltonian $\HH$ is divided into the unperturbed and perturbed parts as 
$\HH = \HH_a + \HH_I$ and the interaction picture is built as
\begin{align}
	A(t) = \hat{U}(t) A_\Hp(t) \hat{U}^{-1}(t)\,,\qquad
	\hat{U}(t) = \mathrm{T} \qty[ \expint{-i}{t}{-\infty}{\HH_I(s)}{s}] \,,
\end{align}
where both pictures are chosen to coincide at $t=-\infty$. The operator equations 
in the interaction picture are
\begin{align} \label{HeisenbergEqInIntPic}
	i\frac{d}{dt} a(t) = \qty[ a(t),\, \HH_a(t)] \,,\qquad
	i\frac{d}{dt} \at(t) = \qty[ \at(t),\, \HH_a(t)]\,,
\end{align}
Except for the fact that it should be bilinear in $a$-operators, the choice of $\HH_a$ 
is non-trivial and essential in nonequilibrium TFD. Although
$\HH$ has no tilde-non-tilde cross-term, such as $a \at$, 
$\HH_a$ contains cross-terms as a result of temporal changes in the thermal situation
and renormalization. For consistency of Eq.~(\ref{HeisenbergEqInIntPic}) under the tilde
conjugation, $\HH_a$ should have the tilde conjugate property of
\begin{align} \label{HaCond1}
	\qty(\HH_a)^\sim = -\HH_a \,.
\end{align}
To obtain Feynman diagrammatic method without relying on
 The Gell-Mann--Low formula, the relation 
$\bra0 \hat{U}^{-1}(\infty) = \bra0$ is necessary \cite{Evans}. 
This is proven in the $\alpha=1$ representation because the relations
 $\bra0 \HH_I(t) = 0$ and 
\begin{align} \label{HaCond2}
	\bra0 \HH_a(t) = 0 \,,
\end{align}
thus hold as in Ref.~\cite{AnnPhys331}.
Equations~\eref{HaCond1} and \eref{HaCond2} are general and critical properties
 of $\HH_a$. 

\section{$4\times4$ formulation}\label{sec-44Formulation}

Herein, we study a general form of $\hat{H}_a$ for a single mode, 
supposing that a BEC exists and the global phase symmetry
is spontaneously broken. At zero temperature, the unperturbed Hamiltonian
involves the symmetry breaking terms such as $aa$ and is diagonalized by 
the usual $2\times2$ Bogoliubov transformation to mix $a$ and $a^\d$ 
\cite{Bogoliubov, deGennes}.
In  finite-temperature and nonequilibrium TFD, we also need the $2\times2$ 
thermal Bogoliubov transformation to mix the tilde and non-tilde operators, as reviewed
in the previous section. Thus, a general $\hat{H}_a$ is a linear combination of 
symmetry-breaking terms, $aa$, $a^\d a^\d$, $\at\at$, $\at^\d \at^\d$, $a^\d\at$, 
and $\at^\d a$,
in addition to the symmetry-preserving terms, $a^\d a$, $\at^\d \at$, $a^\d\at^\d$, 
and $a \at$, and the corresponding transformation has a size of $4\times4$ \cite{Elmfors}.
We treated such symmetry breaking systems in TFD, restricting ourselves to
conduct the two $2\times 2$ transformations sequentially \cite{AnnPhys325, AnnPhys326}.
We discuss the most general form of $\hat{H}_a$ without such a restriction below.

\subsection{$4\times 4$ transformation}

First, we consider a general $4\times 4$ transformation, connecting $a$-operators 
to $\xi$-operators. The properties to be respected are 
the commutation relations \eref{ccr} and \eref{ccr2}, the $\alpha = 1$ 
representation \eref{TSC-a}, and the tilde conjugation rules 
\eref{tilde-rule1} -- \eref{tilde-rule3}. Equation~(\ref{vacua-xi}) for 
the thermal vacua $\bra0$ and $\ket0$ is also assumed.
A general linear transformation for $a$ 
and its tilde conjugate are written as follows:
\begin{align}\label{xi2a-1}
	&a = c \xi + s^* \xit + u_1 \xit^\d + u_2^* \xi^\d \,,\\
	&\at = s \xi + c^* \xit + u_1 \xit^\d + u_2^* \xi^\d \,,
\end{align}
with time-dependent complex coefficients $c$, $s$, $u_1$, and $u_2$.
Although all operators and quantities here and below 
depend on time $t$, $t$ is suppressed for readability. 
Then, Eq.~\eref{TSC-a} restricts the forms of $\at^\d$ and $a^\d$ as 
\begin{align}
	\at^\d &= c   \xi + s^*   \xit + u_3   \xit^\d + u_4^*   \xi^\d \,,\\
	a^\d   &= s \xi + c^* \xit + u_4 \xit^\d + u_3^* \xi^\d \,,\label{xi2a-4}
\end{align}
with the additional coefficients $u_3$ and $u_4$. Note that $\xi^\d$ and $\xit^\d$ 
are not Hermitian conjugates of $\xi$ and $\xit$, respectively. The commutation relations 
\eref{ccr} lead to the relations
\begin{align}\label{u3-u1}
	\pmqty{u_3 -u_1 \\ u_4^* - u_2^*} 
	= \frac{1}{\Delta} \pmqty{c & -s \\ -s^* & c^*} \pmqty{1 \\ 0} \,,
\end{align}
with $\Delta = |c|^2 - |s|^2$, where $\Delta\ne0$ is assumed. The most characteristic 
and independent expectations of the bilinear $a$-operators,
 $\ev{a^\d a}{0}$ and $\ev{a a}{0}$, are calculated as
\begin{alignat}{2}
	n &= \ev{a^\d a}{0} &&= s u_2^* + c^*u_1 \,,\\
	m &= \ev{a\; a}{0} &&= cu_2^* +s^* u_1\,, 
\end{alignat}
which is inverted as  
\begin{align}
	\pmqty{u_1 \\ u_2^*} = \frac{1}{\Delta}  \pmqty{c & -s \\ -s^* & c^*} \pmqty{n\\  m} \,.
\end{align}
From Eq.~\eref{u3-u1}, $u_3$ and $u_4$ are expressed as
\begin{align}
	\pmqty{u_3 \\ u_4^*} = \frac{1}{\Delta}  \pmqty{c & -s \\ -s^* & c^*} \pmqty{1+n\\  m} \,.
\end{align}
We rearrange the coefficients as
\begin{align}
	\pmqty{u_1 & u_2^* \\ u_2 & u_1^*} &= \pmqty{n & m \\ m^* & n} \pmqty{c_* & -s_*^* \\ -s_* & c_*^*} \,,\\
	\pmqty{u_3 & u_4^* \\ u_4 & u_3^*} &= \pmqty{1+n & m \\ m^* & 1+n} \pmqty{c_* & -s_*^* \\ -s_* & c_*^*} \,,
\end{align}
with $c_* = c/\Delta$ and $s_*=s/\Delta$.
Then, through simple manipulations to eliminate $u_1$ -- $u_4$, 
Eqs.~\eref{xi2a-1} -- \eref{xi2a-4} are put into a $4\times 4$ matrix form, using
the $4\times 4$ Bogoliubov 
matrix $\Bc$ and the quartet notation, 
\begin{align} \label{thermalBogoiubov}
	a^i = \Bc^{-1, ij} \xi^j\,,\qquad
	a^i = \pmqty{a \\ \at \\ \at^\d \\ a^\d}^i \,,\qquad
	\xi^i = \pmqty{\xi \\ \xit \\ \xit^\d \\ \xi^\d}^i \,.
\end{align}
Here, summations of repeated superscripts are implicit. Defining 
the block matrices,
\begin{align}\label{def-WandN}
	W = \pmqty{c_*^* & -s_*^* \\ -s_* & c_*} \,,\qquad
	W^{-1} = \pmqty{c & s^* \\ s & c^*} \,,\qquad 
	N = \pmqty{n & m \\ m^* & n}\,,
\end{align}
the unity matrix $I$, and zero matrix $O$, we have the following 
compact expressions,
\begin{align}\label{BcWN}
	\Bc^{-1} = \bmqty{I & N \\ I & I+N} \bmqty{W^{-1} & O \\ O & W^\d}\,,\qquad
	\Bc = \bmqty{W & O \\ O & W^{\d,-1}} \bmqty{I+N & -N \\ -I & I} \,,
\end{align}
where a block matrix is denoted by $\bmqty{\quad}$ to distinguish from a simple 
matrix $\pmqty{\quad}$. 
The bar-conjugate of the quartet is also introduced as
\begin{align}
	\ab^i = \pmqty{a^\d &  \at^\d & -\at & -a }^i \,,\qquad
	\xib^i = \pmqty{\xi^\d &  \xit^\d & -\xit & -\xi}^i \,.
\end{align}
The transformation of the bar-conjugate is $\ab = \xib \Bc$.

\subsection{General form of $\HH_a$}

Next, we derive a general form of $\HH_a$ 
under the conditions \eref{HaCond1} and \eref{HaCond2}. 
Although the parameters in $\HH_a$ are time-dependent in general nonequilibrium cases, 
let us consider for a moment the case 
in which all parameters are time-independent.
The condition \eref{HaCond1} allows
\begin{multline}
	\HH_a =
		 (d_1-id_2) a^\d a - (d_1+id_2) \at^\d \at 
		 +i d_3 \at a + id_4 \at^\d a^\d 
		 \\
		 +c_1 a^\d a^\d - c_1^*\at^\d \at^\d 
		 +c_2 \at \at - c_2^* a^\d a^\d 
		 +c_3 \at a^\d 	-c_3^* a \at^\d	
		\,,
\end{multline}
where $d_i$ and $c_i$ are real and complex constants, respectively. 
Then, Eq.~\eref{HaCond2} with Eq.~\eref{TSC-a} applies the relations,
\begin{align}
	2d_2 = d_3 + d_4 \,, \qquad
	c_1 + c_2 + c_3 = 0\,.	
\end{align}
For later discussions, we introduce three real parameters, i.e., $\Lc$, $\gamma$, and $\kappa$,
and two complex parameters, $\eta$ and $\Mc$, instead of $d_i$ and $c_i$:
\begin{align}
	d_1 = \Lc \,,\qquad 
	d_2 = \gamma \,,\qquad
	d_3 = \gamma+\kappa \,,\qquad
	d_4 = \gamma-\kappa \,,\\
	c_1 = \frac{\Mc-\eta}{2} \,,\qquad
	c_2 = -\frac{\Mc+\eta}{2} \,,\qquad
	c_3 = \eta \,.
\end{align} 
We then obtain
\begin{align}
	\HH_a &= \frac12\;\ab^i
	\pmqty{\begin{array}{cc|cc}
		\Lc - i\gamma  & \eta & i(\gamma-\kappa)& \Mc - \eta \\
		-\eta^* & -\Lc - i\gamma  & -\Mc^* + \eta^* & i(\gamma - \kappa) \\
		\hline
		-i(\gamma+\kappa) & \Mc + \eta & \Lc +i\gamma & -\eta \\
		-\Mc^* - \eta^* & -i(\gamma+\kappa) & \eta^* & -\Lc + i\gamma 
	\end{array}}^{ij}
	a^j \label{eq:Ha}\\
	&=\frac12\;\ab^i
	\bmqty{\Hc^{11} & \Hc^{12} \\ \Hc^{21} & \Hc^{22}}^{ij}
	a^j\,. \label{def-Hcij}
\end{align}
The last equality aims to define the values of $\Hc$.

Thus far, we have argued the general forms of $\Bc$ and $\HH_a$ separately.
Both
arguments must be combined for consistency, as described below.
Although the thermal vacua $\ket{0}$ and $\bra{0}$ are time-independent, they 
are specified by the time-dependent values of $\xi(t)$, as shown in Eq.~\eref{vacua-xi}.
This requires that $\HH_a$ be diagonal in terms of $\xi$-operators
when $\Bc$ is time-independent.
When Eqs.~(\ref{thermalBogoiubov})--(\ref{BcWN}) are substituted into
Eq.~\eref{def-Hcij}, $\HH_a$ is expressed in terms of the $\xi$-operator as 
\begin{align}
	\HH_a 
	= \frac12\;\xib^i 
	\bmqty{WTW^{-1} & W \qty(\Hc^{12} +TN-NT^\d)W^\d \\ O & \qty(WTW^{-1})^\d}^{ij} \xi^j \,,
	\label{Hx-with-xi}
\end{align}
where $T$ is defined by
\begin{align}\label{eq:def-T}
	T = \Hc^{11} + \Hc^{12} = \Hc^{21} + \Hc^{22} 
	= \pmqty{ \Lc - i\kappa & \Mc \\ -\Mc^* &-\Lc-i\kappa}\,.
\end{align}
The diagonal condition of $\HH_a$ in Eq.~\eref{def-Hcij} is reduced to
the two conditions $\Hc^{12} +TN-NT^\d = O$ and diagonal $WTW^{-1}$.
Explicitly, the first condition is 
\begin{align}
	&\Hc^{12} +TN-NT^\d \notag \\
	&\quad = \pmqty{
		i(\gamma-\kappa) +2i\kappa n -2i\Im\qty[\Mc^* m] & 
		\Mc-\eta + 2\Mc n +2(\Lc-i\kappa)m\\
		-\Mc^*+\eta^* - 2\Mc^* n -2(\Lc+i\kappa)m^* & 
		i(\gamma-\kappa)+ 2i\kappa n -2i\Im\qty[\Mc^* m]
	} = O\,.
\end{align}
The parameters $\gamma$ and $\eta$ are fixed as
\begin{align}
	\gamma = (1+2n)\kappa + 2\Im\qty[\Mc^* m] \,,\qquad
	\eta = (1+2n)\Mc + 2(\Lc-i\kappa)m \,.
	\label{eq:gammaeta}
\end{align}
The second condition is the eigenvalue problem of $T$. This is 
a familiar BdG equation for $T$ without $\kappa$ \cite{Fetter} and
appears in the system of a dissipative condensate for $T$ with a nonvanishing 
$\kappa$ \cite{Chen}. 

Because $T$ is non-Hermitian, we set the right and left eigenfunctions with eigenvalue $E$ as $y$ and $y_*^\d$, respectively:
\begin{align}\label{lrBdG}
	T y = E y \,,\qquad
	y_*^\d T = E y_*^\d\,.
\end{align}
We proceed under the presupposition that the real parts of both $E$ and $y_*^\d y$ are non-zero
such that the eigenfunctions can be normalized as $y_*^\d y = 1$. 
Because of the symmetry $\sigma_1 T \sigma_1 = -T^*\,$ with the Pauli matrix 
$\sigma_1$, $-E^*$ is also an eigenvalue and
the corresponding right and left eigenfunctions are $z=\sigma_1 y^*$ and $z_*^\d = z_*^t \sigma_1$, respectively, which are normalized by $z_*^\d z = -1$.
Combining them with the trivial orthogonality because of $E\ne-E^*$, we have
\begin{align} \label{tdcs1}
	\pmqty{y_*^\d \\ - z_*^\d} \; \pmqty{y & z} = I\,.
\end{align}
Therefore, the choice of $W^{-1}$ and $W$ as
\begin{align}
	W^{-1} = \pmqty{y & z}\,,\qquad
	W = \pmqty{y_*^\d \\ z_*^\d}\,,
\end{align}
fulfills the second condition,
\begin{align}
	\label{eigensystemT}
	WTW^{-1}  =  \pmqty{E & 0 \\ 0 &-E^*} \,.
\end{align}
In the present single mode case, 
we can explicitly solve the eigenvalue problem and achieve
\begin{align}\label{eq:defEyy}
	E  = \omega-i\kappa \,,\qquad
	y  = \frac1{\sqrt{2\omega}} \pmqty{ \sqrt{\Lc+\omega} \\[1 mm] -\sqrt{\Lc-\omega}} 
\quad \quad
	y_*= \frac1{\sqrt{2\omega}} \pmqty{ \sqrt{\Lc+\omega} \\[1 mm]  \sqrt{\Lc-\omega}}
\end{align}
with $\omega = \sqrt{\Lc^2 - |\Mc|^2}$.

In general, $\omega$ can be either complex or zero, and $y_*^\d y$ then vanishes.
We consider real and non-zero $\omega$ solely in this paper.
According to Eqs.~\eref{Hx-with-xi} and \eref{eigensystemT}, we successfully diagonalize $\HH_a$ as
\begin{align}
	\HH_a = \frac12\;\xib^i 
	\pmqty{\begin{array}{cc|cc}
			E &&&  \\
		&  -E^* &&  \\ \hline
		&&  E^* & \\
		&&& -E  
	\end{array}}^{ij} \xi^j 
	= E\xi^\d \xi -E^*\xit^\d \xit \,, 
	\label{diagonalizedHx}
\end{align}
and we have well-defined thermal vacua for arbitrary $t$:
$\bra0 \xi^\d(t) = \bra0 \xit^\d(t) = 0\,,\quad	\xi(t)\ket0 = \xit(t)\ket0 = 0\,$.

Now, let us take account of the time-dependence of $\Bc$. 
The Hamiltonian $\HH_a$ in \eref{diagonalizedHx} is the time-evolution generator of 
$\xi$-operators, but not $a$-operators. To deal with the time dependence of $\Bc$,
or of $N$ and $W$ in Eq.~\eref{BcWN}, we introduce two extra generators, 
$\hat{Q}$ and $\hat{R}$:
\begin{align}
	\hat{Q}&= -\frac{i}{2} \xib^i \bmqty{O & W\dot{N}W^{\d} \\ O & O}^{ij} \xi^j \,,
	\qquad\qquad
	\hat{R}&= \frac{i}{2} \xib^i \bmqty{\dot{W}W^{-1} & O \\ O & \dot{W}^{\d,-1}W^\d}^{ij} \xi^j\,.
\end{align}
Their contributions should be compensated somehow to keep stationary thermal vacua.
In the following, we take separate approaches for $\hat{Q}$ and $\hat{R}$\,.

As for $\hat{Q}$, we treat it as a counter term, which is well established 
in nonequilibrium TFD \cite{AIP} and is labeled a thermal counter term,  
meaning that $\HH_a$ in Eq.~\eref{diagonalizedHx} [see also Eq.~\eref{Hx-with-xi}]
is shifted, i.e.,
\begin{align}
	\HH_a = \frac12\;\xib^i 
	\bmqty{WTW^{-1} & O \\ O & \qty(WTW^{-1})^\d}^{ij} \xi^j -  \hat{Q}\,.
\label{eq:genHa}
\end{align}
This shift corresponds to the simple parameter shifts, 
$\gamma\to \gamma+\nd$ and $\eta\to\eta -i\md$\, and therefore,  
Eqs.~\eref{HaCond1} and \eref{HaCond2} remain as they are.

Next, for $\hat{R}$\, we follow the treatment proposed by 
Matsumoto and Sakamoto \cite{Matsumoto2001}, 
which we refer to as a TDCS method. 
With this method, $W$ is determined using the eigenvalue problem of $T$ [see Eq.~\eref{eigensystemT}]  
only initially, and thereafter is subject to the following time-dependent 
BdG equations:
\begin{align}\label{TDBdG-W}
	 i\frac{d}{dt} W^{-1} = T W^{-1}\,,\qquad
	-i\frac{d}{dt} W = WT\,.\qquad
\end{align}
Here, $W^{-1}$ remains an inverse of $W$ for any $t$ because $i\frac{d}{dt} \qty{ WW^{-1}} = WTW^{-1} - WTW^{-1} = 0 \,$.
The introduction of $W$ in this way enables us to take in its time dependence,
keeping  the stationary thermal vacua, and without adding $\Hat{R}$ as a counter term. 
This is because $\xi^i(t)$ is 
independent of $t$, as
\begin{align}
	i\frac{d}{dt} \xi^i 
	= i\dot{\Bc}^{ij}\, a^j + \Bc^{ij}\qty[a^j ,\, \HH_a]  = 0\,,
\label{eq:TDSC-dtxi}
\end{align}
and the thermal vacua that $\xi^i(0)$ annihilate are annihilated
by $\xi^i(t)$ for any $t$. In other words, the original time-dependent 
diagonal factor of $\xi^i(t)$ 
is moved to $W(t)$ in the TDCS.
For convenience, we write the construction of the TDCS in terms of $c_{(*)}$ and $s_{(*)}$.
The initial conditions of $c_{(*)}$ and $s_{(*)}$ are given by the following eigenvalue problem:
\begin{align}
	&T(0) \pmqty{c(0) \\ s(0) } = E \pmqty{c(0) \\ s(0) },\quad
	T^\d(0) \pmqty{c_*(0) \\- s_*(0) } = E^* \pmqty{c_*(0) \\ -s_*(0) }\,, \\
	&c_*^*(0) c(0) - s_*^*(0)s(0) = 1 \,,
\end{align}
and evolve as
\begin{align}\label{TDBdG-y}
	i\frac{d}{dt} \pmqty{c(t) \\ s(t) } = T(t)\pmqty{c(t) \\ s(t) } \quad \quad \quad
	i\frac{d}{dt} \pmqty{c_*(t) \\ -s_*(t) } = T^\d(t)\pmqty{c_*(t) \\ -s_*(t) }.
\end{align}
Then, $W$ is expressed as in Eq.~\eref{def-WandN}.

Summarizing this section, we have derived the general $4\times4$ transformation 
$\Bc$ and the unperturbed hat Hamiltonian $\HH_a$ in nonequilibrium TFD. 
Although the time dependence of $N$ is incorporated as a thermal counter term, 
we deal with that of $W$ by setting up a TDCS.
In $\HH_a$, the three real parameters, $\Lc$, $\kappa$, and $\nd$, 
and two complex parameters, $\Mc$ and $\md$, are independent. 
They will be determined through the renormalization conditions
in Sec.~\ref{Sec:renormalization}. 
The parameters of $W$ in $\Bc$, $c_{(*)}$, and $s_{(*)}$ are calculated 
from the TDCS.

Here we compare the present $4\times4$ formulation with our previous 
version \cite{AnnPhys325,AnnPhys326}. In conclusion, although the matrix $\Bc$ 
is identical in both formulations, the form of $\HH_a$ in the present formulation
 is more general in the sense that $\HH_a$ in this study involves 
more parameters than those in the previous formulation. 

\section{Model of quantum field theory}\label{sec-ModelQFT}

The $4\times 4$ formulation for a single mode
 in Sec.~\ref{sec-44Formulation} is applied to
a system based on quantum field theory. For this, we consider the Hamiltonian for a bosonic field $\psi$ with a repulsive, contact-type self-interaction: 
and without an external potential
in the Heisenberg picture, 
\begin{align}
H=\intx \left[\psi^\d_H(x) \left\{-\frac{\nabla^2}{2m}-\mu \right\}
\psi_H(x) + \frac{g}{2}\psi^{\d}_H(x)\psi_H^{\d}(x) \psi_H(x)\psi_H(x)  \right]\,,
\label{eq:H}
\end{align}
where $x=(\bx, t)$ and $m$, $\mu$, and $g\, (>0)$ represent the mass of the particle, 
the chemical potential, and the repulsive coupling constant, respectively. 
The field operators, $\psi_H(x)$ and $\psi^\d_H(x)$,
 satisfy the canonical commutation relations
 \begin{align}
 &[\psi_H(x), \psi^\d_H(x')]|_{t=t'}=\delta(\bm{x}-\bm{x'}),\notag\\
 &[\psi_H(x), \psi_H(x')]|_{t=t'}=[\psi^\d_H(x), \psi_H^\d(x')]|_{t=t'}=0 \,.
 \end{align}
The system has a symmetry of a global phase transformation, $\psi_H(x)\,\rightarrow
\, \psi_H(x)\, e^{i\theta}$ with a constant $\theta$, 
in addition to symmetries of a spatial translation and rotation. 
We assume throughout this paper 
that a stationary and homogeneous condensate exists, which means that the 
global phase symmetry is spontaneously broken but the translational and rotational symmetries
are preserved. Then, based on quantum field theory at zero temperature, 
$\psi_H(x)$ is divided into a constant $c$-number $v$, called the order parameter, 
and a new field operator, $\phi_H(x)$, as follows:
\begin{align}
\psi_H(x)  = v + \phi_H(x) \,,\qquad
v = \expvac{\psi_H} = \sqrt{n_c}\,,
\end{align}
where $n_c$ is the density of a condensed particle.
The total Hamiltonian is expressed as the sum of the free part $H_0$, which is either linear 
or bilinear in $\phi$, and the interaction part $H_{\mathrm {int}}$ for the remainder,
\begin{align}
H &= H_0+H_{\mathrm {int}}\,,\label{original-H}\\
H_0
&= \int\dd[3]{x} \qty(\phi_H + \phi^\d_H)\qty(- \mu + gn_c )\sqrt{gn_c}
\notag \\
&\quad + \frac12\int\dd[3]{x} \pmqty{\phi^\d_H & -\phi_H}\pmqty{-\frac{\nabla^2}{2m}
 - \mu + 2gn_c & gn_c \\ -gn_c & \frac{\nabla^2}{2m}+\mu-2gn_c}
\pmqty{\phi_H \\ \phi^\d_H}, \label{eq:H0}\\
H_{\mathrm {int}}&=\int\dd[3]{x} \qty[g\sqrt{n_c} \qty(\phi_H^\d\phi_H^\d\phi_H
 + \phi^\d_H\phi_H\phi_H)
 + \frac{g}{2}\phi_H^\d\phi_H^\d\phi_H\phi_H]\,.
\end{align}
Note that the $\phi$-linear terms in $H_0$ vanish under the 
condition $\expvac{\phi_H(x)}=0$
for any $t$, which leads to $\mu= g n_c$ in the tree approximation.
  
Because the system is homogeneous, it is useful to expand $\phi_H(x)$ in momentum eigenfunctions,
\begin{align}
&\phi_H(x) =  \frac{1}{\sqrt{V}}\sum_{\bq}\; a_{H,\bq}(t)e^{i\bq\cdot \bx}\,,
\end{align}
where we introduce periodic boundary conditions, and $V$ is the volume of the box. Then, 
the free Hamiltonian (\ref{eq:H0}) in a tree approximation becomes the following:
\begin{align}
H_0 = \frac{1}{2}\sum_{\bq}\pmqty{a^\d_{H,\bq} & -a_{H,-\bq}}
\pmqty{\ve_q+gn_c & gn_c \\ -gn_c & -\ve_q-gn_c}\pmqty{a_{H,\bq} \\ a^\d_{H,\bq}}\,,
\end{align}
where $\ve_q=q^2/2m\,$. The interaction Hamiltonian $H_{\rm int}$ is given in terms
of the $a$-operators as
\begin{align}
H_{\rm int}&=g\sqrt{\frac{n_c}{V}} \sum_{\bq_1,\bq_2,\bq_3}
\qty(a^\d_{H,\bq_1} a^\d_{H,\bq_2}a_{H,\bq_3}\delta_{\bq_1+\bq_2, \bq_3}
+a^\d_{H,\bq_1} a_{H,\bq_2}a_{H,\bq_3}\delta_{\bq_1, \bq_2+\bq_3})
\notag \\
&\qquad 
+ \frac{g}{2V} \sum_{\bq_1,\bq_2,\bq_3,\bq_4} 
a^\d_{H,\bq_1} a^\d_{H,\bq_2}a_{H,\bq_3}a_{H,\bq_4}
\delta_{\bq_1+\bq_2, \bq_3+\bq_4}\,.
\end{align}

\section{Hamiltonians of nonequilibrium TFD for quantum field system}\label{sec-NETFDQFT}

We turn to the formulation of nonequilibrium TFD for the system above.
As in Sec.~\ref{sec-44Formulation}, the quartet notation is introduced. 
The quartet of the $a$-operators in the interaction picture is
related to the quartet of the
 $\xi$-operators as follows:
\begin{align}
&a^i_{\bq}(t)=\Bc^{-1, ij}_{\bq}(t)\xi^j_{\bq}(t),\label{Bogoliubov}\quad 
\ab^i_{\bq}(t)=\xib^j_{\bq}(t)\Bc^{ji}_{\bq}(t),\\
&a^i_{\bq}(t)=\pmqty{a_{\bq} \\ \at_{-\bq} \\ \at^\d_{\bq} \\ a^\d_{-\bq}}^i_t \,,\qquad
 \ab^i_{\bq}(t) = \pmqty{a_\bq^\d & \at_{-\bq}^\d & -\at_{\bq} & -a_{-\bq}}^i_t \,,\\
&\xi^i_{\bq}(t)=\pmqty{\xi_{\bq} \\ \xit_{-\bq} \\ \xit^\d_{\bq} \\ \xi^\d_{-\bq}}^i_t\,, 
\qquad
\xib^i_{\bq}(t) = \pmqty{\xi_\bq^\d & \xit_{-\bq}^\d & -\xit_{\bq} & -\xi_{-\bq}}^i_t\,.
\end{align}
The $4\times 4$ matrix $\Bc_\bq$ is
\begin{align}
& \Bc^{-1}_{\bq}(t)=\bmqty{I & N_{\bq} \\ I & I+N_{\bq}}_t
\bmqty{W^{-1}_{\bq} & O 
\\ O & W^\d_{\bq}}_t\,, \quad 
\Bc_{\bq}(t)=\bmqty{W_{\bq} & O \\ O & W^{\d,-1}_{\bq}}_t
\bmqty{I+N_{\bq} & -N_{\bq} \\ -I & I}_t\,,
\label{eq:Bc}
\\
&  W_{\bq}(t) 
= \pmqty{c^{*}_{*\bq} & -s^{*}_{*\bq} \\ -s_{*\bq} & c_{*\bq}}_t\,,\qquad
 W^{-1}_{\bq}(t) 
= \pmqty{c_{\bq} & s^{*}_{\bq} \\ s_{\bq} & c^*_{\bq}}_t\,,\qquad
N_{\bq}(t) = \pmqty{n_{\bq} & m_{\bq} \\ m^{*}_{\bq} & n_{\bq}}_t\,,
\label{eq:WN}
\end{align}
where $n_{\bq}(t)$ and $m_{\bq}(t)$ represent time-dependent 
normal and anomalous condensate number distributions, respectively,
\begin{align}
 n_{\bq}(t)\delta_{\bq \bq'}= \expvac{a^\d_{\bq}(t)a_{\bq'}(t)}
\,,\qquad m_{\bq}(t)\delta_{\bq \bq'}= \expvac{a_{\bq}(t)a_{-\bq'}(t)}\,.
\end{align}
The canonical commutation relations are
\begin{align}
& [a^i_{\bq}(t), \ab^j_{\bq'}(t)]=\delta_{ij}\delta_{\bq\bq'}\,,
\quad [a^i_{\bq}(t), a^j_{\bq'}(t)]= [\ab^i_{\bq}(t), \ab^j_{\bq'}(t)]=0\,,\\
& [\xi^i_{\bq}(t), \xib^j_{\bq'}(t)]=\delta_{ij}\delta_{\bq\bq'}\,,
\quad [\xi^i_{\bq}(t), \xi^j_{\bq'}(t)]=[\xib^i_{\bq}(t), \xib^j_{\bq'}(t)]=0
\, .
\end{align}

The total hat Hamiltonian of the system under consideration is
\begin{align}\label{H1}
\HH =\HH_0+\HH_{\mathrm {int}} \,.
\end{align}
The hat free Hamiltonian is expressed in the quartet notation as
\begin{align}
\hat{H}_0 = \frac{1}{2}\sum_{\bq}\;\ab^i_\bq
\pmqty{\begin{array}{cc|cc}
	\ve_q+gn_c  &&& gn_c  \\
	& -\ve_q-gn_c & -gn_c & \\
	\hline
	& gn_c & \ve_q+gn_c & \\
	-gn_c  &&& -\ve_q-gn_c  
	\end{array}}^{ij}
a^j_\bq.
\end{align}

Because the self-consistent
renormalization conditions determine the parameters in ${\hat H}_a$ or yield 
time evolution equations for them, for example, the equations for $n_{q}(t)$ and $m_{q}(t)$, the choice of unperturbed hat Hamiltonian ${\hat H}_a$
is a crucial step in nonequilibrium TFD.
The general form of ${\hat H}_a$ for a single mode in Eq.~(\ref{eq:genHa}) 
can be transferred to the present case with the addition of the momentum indices.
See also Eqs.~(\ref{eq:Ha}) and (\ref{eq:gammaeta}). 
For the momentum index $\bq$, all c-number parameters appearing in ${\hat H}_a$, 
and $\Bc$ in Eqs.~(\ref{eq:Bc}) and (\ref{eq:WN}), 
are even functions of $\bq$. Furthermore, if the initial condition is isotropic,
they depend only on $q = |\bq|$ for any $t$. We restrict our discussions
below, and denote the momentum indices of the parameters simply by $q$.
Thus, ${\hat H}_a$, which generates the temporal evolution of $a^i_\bq(t)$,
\begin{align}
i\frac{d}{dt} a^i_{\bq}=[a^i_{\bq}, \HH_a] \,,
\end{align}
is given by 
\begin{align}
\HH_a&=\HH_\xi-\HQ \,, \label{eq:Haq} \\
\HH_\xi &= \frac12 \sum_{q}\;
\ab^i_{\bq}
\pmqty{\begin{array}{cc|cc}
	\Lc_q - i\gamma_q  & \eta_q & i(\gamma_q-\kappa_q)& \Mc_q - \eta_q \\
	-\eta_q^* & -\Lc_q - i\gamma_q  & -\Mc_q^* + \eta_q^* & i(\gamma_q - \kappa_q) \\
	\hline
	-i(\gamma_q+\kappa_q) & \Mc_q + \eta_q & \Lc_q +i\gamma_q & -\eta_q \\
	-\Mc_q^* - \eta_q^* & -i(\gamma_q+\kappa_q) & \eta_q^* & -\Lc_q + i\gamma_q 
	\end{array}}^{ij}
a^j_{\bq} \\
& =
 \frac12 \sum_{q} \xib^i_{\bq}  
	\bmqty{W_qT_qW^{-1}_q & O \\ O & \qty(W_qT_qW^{-1}_q)^\d}^{ij} \xi_{\bq}^j
\,, \label{eq:Hxiq} \\
 T_q &=\pmqty{ \Lc_q - i\kappa_q & \Mc_q \\ -\Mc^*_q &-\Lc_q-i\kappa_q}\,,\\
\HQ&=-\frac{i}{2}\sum_{\bq}\ab^i_{\bq}
\bmqty{-\Nd_q & \Nd_q \\ -\Nd_q & \Nd_q}^{ij}a^j_{\bq} 
=-\frac{i}{2}\sum_{\bq}\xib^i_{\bq}\bmqty{O & W_q\Nd_q W_q^\d \\ O & O}^{ij}\xi^j_{\bq}\,.
\label{eq:HQq}
\end{align}
The two parameters, $\gamma_q$ and $\eta_q$, 
are given by the other parameters, corresponding to Eq.~(\ref{eq:gammaeta}), 
\begin{align}
& \gamma_q = (1+2n_q) \kappa_q + 2\Im\qty(\Mc_q^* m_q) \,,\label{def_gamma}\\
&\eta_q  = (1+2n_q) \Mc_q + 2\qty(\Lc_q -i\kappa_q)m_q \,.\label{def_eta}
\end{align}
Note that $W_q$, which represents the usual Bogoliubov transformation in equilibrium, 
depends on time in nonequilibrium, although a stationary condensate is assumed. 
The time-dependence of $W$ is not dealt with as a counter term, but by introducing the TDCS, 
as described in Sec.~\ref{sec-44Formulation}. Namely, the time-dependent BdG equations
in Eq.~(\ref{TDBdG-W}) are extended to
\begin{align}\label{TDBdG-Wq}
	i\frac{d}{dt} W^{-1}_q = T_q W^{-1}_q\,,\qquad
	-i\frac{d}{dt} W_q = W_qT_q\,.
\end{align} 

We define the counter Hamiltonian $\delta\HH$ as
\begin{align}
\delta\HH&=\HH_\xi-\HH_0 \label{eq:def-deltaHH}\\
&=\frac{1}{2}\sum_{\bq}
\ab^i_\bq
\pmqty{\begin{array}{cc|cc}
	\delta\Lc_q - i\gamma_q  & \eta_q & i(\gamma_q-\kappa_q)& \delta\Mc_q - \eta_q \\
	-\eta_q^* & -\delta\Lc_q - i\gamma_q  & -\delta\Mc_q^* + \eta_q^* & i(\gamma_q - \kappa_q) \\
	\hline
	-i(\gamma_q+\kappa_q) & \delta\Mc_q + \eta_q & \delta\Lc_q +i\gamma_q & -\eta_q \\
	-\delta\Mc_q^* - \eta_q^* & -i(\gamma_q+\kappa_q) & \eta_q^* & -\delta\Lc_q + i\gamma_q 
	\end{array}}^{ij}
a^j_\bq \,,\label{eq:dH-a}\\
&=\frac{1}{2}\sum_{\bq}\xib^i_\bq\bmqty{
	W_q \delta T_qW_q^{-1} & W_q \Xc_q W^\d_q \\ 
	O & W^{-1,\d}_q \delta T_q^\d W^\d_q
	}\xi^j_\bq \,,\label{eq:dH-xi}
\end{align}
where
\begin{align}
& \delta\Lc_q = \Lc_q - (\ve_q+gn_c)\,,\qquad
\delta\Mc_q = \Mc_q - gn_c\,, \label{eq:defdeltaL}\\
& \delta T_q= 
\pmqty{ \delta \Lc_q - i\kappa_q & \delta \Mc_q \\ -\delta
\Mc^*_q &-\delta \Lc_q-i\kappa_q}\,,\label{eq:defdeltaT}\\
& \Xc_q=\pmqty{2i\Im(m_qgn_c) & -(1+2n_q)gn_c-2m_q(\ve_q+gn_c) \\
	(i+2n_q)gn_c+2m_q^*(\ve_q+gn_c) & 2i\Im(m_qgn_c)}\,.
\label{eq:defXc}
\end{align}

\section{Feynman method and self-energy}\label{sec-FeynmanSE}

The interaction picture of nonequilibrium TFD is formulated using
the unperturbed hat Hamiltonian ${\hat H}_a$ in Eqs.~(\ref{eq:Haq}) --
 (\ref{eq:HQq}) using Eq.~\eref{eq:def-deltaHH} and perturbed 
Hamiltonian ${\hat H}_I$,
\begin{align}
&\HH=\HH_a+\HH_I\,, \\
&\HH_I= \HH_{\rm int}-\delta\HH+\HQ \,.
\end{align}
We take $\Bc$ in the $\alpha=1$ representation \cite{AIP},
and can thus make use of the Feynman diagram method in calculating the 
full propagators \cite{Evans}. The full propagator of the $a$-operator, 
defined by $G^{ij}_\bq(t_1,t_2) \delta_{\bq\bq'}
=-i\expvac{\mathrm{T}[a^{i}_{H, \bq}(t_1)\ab^{j}_{H,\bq'}(t_2)]}$, 
can be obtained by calculating the formula, 
\begin{align}
G^{ij}_\bq(t_1,t_2)\delta_{\bq\bq'}
= -i\expvac{\mathrm{T}[{\hat S} a^{i}_{\bq}(t_1)\ab^{j}_{\bq'}(t_2)]}\,,
\label{eq:SMatrix}
\end{align} 
where ${\hat S}={\hat U}(\infty)$ is the S-matrix operator corresponding to the 
perturbed Hamiltonian $\HH_I$ (see the sentences below Eq.~\eref{HaCond1})\,. 
The unperturbed propagator of $a$-operator is
\begin{align}
\Delta^{ij}_\bq(t_1,t_2)\delta_{\bq\bq'}&=-i\expvac{\mathrm{T}[a^{i}_{\bq}(t_1)
\ab^{j}_{\bq'}(t_2)]} \,.
\end{align}
Now that the Feynman diagram method is available,
the self-energy is denoted by $\Sigma_\bq^{ij}$ and the Dyson equation reads as
\begin{align}
G^{ij}_\bq(t_1,t_2)=\Delta^{ij}_{\bq}(t_1,t_2)
+\int\! ds_1ds_2\,\Delta_\bq^{ii'}(t_1,s_1)\Sigma_\bq^{i'j'}
(s_1,s_2)G_\bq^{j'j}(s_2,t_2)\,.\label{Dyson}
\end{align}

Similar arguments can be repeated for the full and unperturbed
propagators of the $\xi$-operator, which is defined by
\begin{align}
&g^{ij}_\bq(t_1,t_2)\delta_{\bq\bq'}
=-i\expvac{\mathrm{T}[\xi^{i}_{H,\bq}(t_1)\xib^{j}_{H,\bq'}(t_2)]}\label{xi-FullPropagator}\,,\\
&d^{ij}_\bq(t_1,t_2)\delta_{\bq\bq'}=-i\expvac{\mathrm{T}[\xi^{i}_{\bq}(t_1)\xib^{j}_{\bq'}(t_2)]}
\,.
\end{align}
The latter is explicitly 
\begin{align}
d^{ij}_{\bq}(t_1,t_2) = \bmqty{ 
	-i\theta(t_1-t_2) I&0\\
	O& i\theta(t_2-t_1)I }^{ij}\,.
\end{align}
We should remark that $\xi^i(t)$ is independent of $t$, as in Eq.~\eref{eq:TDSC-dtxi}, whereas
$t$ is necessary for arranging the operators in order of time.
The Dyson equation is
\begin{align}
g^{ij}_\bq(t_1,t_2)=d^{ij}_\bq(t_1,t_2)+\int\! ds_1ds_2\,d_\bq^{ii'}(t_1,s_1)
S_\bq^{i'j'}(s_1,s_2)g_\bq^{j'j}(s_2,t_2)\,,
\end{align}
where $S_q^{ij}$ is the self-energy of the $\xi$-operator.

The propagators $G^{ij}_\bq$ and $\Delta^{ij}_\bq$ are related to 
$g^{ij}_\bq$ and $d^{ij}_\bq$ 
through the Bogoliubov transformation (\ref{Bogoliubov}) as
\begin{align}
& g^{ij}_\bq(t_1,t_2) = \Bc^{ii'}_q(t_1) G^{i'j'}_\bq (t_1,t_2) \Bc^{-1, j'j}_q
(t_2)\,,\\
& d^{ij}_\bq(t_1,t_2) = \Bc^{ii'}_q(t_1) \Delta^{i'j'}_\bq(t_1,t_2) \Bc^{-1, j'j}_q(t_2)
\,.
\end{align}
From these relations and the Dyson equations above, we obtain
\begin{align}
S^{ij}_{\bq}(t_1,t_2) = \Bc^{ii'}_{q}(t_1) \Sigma^{i'j'}_{\bq}
(t_1,t_2)\Bc^{-1,j'j}_{q}(t_2) \,.\label{xi-selfenergy}
\end{align}

The lowest-order contributions of the thermal counter term ${\hat Q}$ in Eq.~\eref{eq:HQq}
and the counter Hamiltonian $\delta \HH$ in Eq.~\eref{eq:dH-xi}
to the self-energies are calculated as
\begin{align}
&S^{ij}_{Q, \bq}(t_1,t_2)
   = -i\bmqty{  O& W_q\dot{N}_q W_q^\d \\ O & O}_{t_1}^{ij} \delta(t_1-t_2),\\
&S^{ij}_{\delta H, \bq}(t_1,t_2)=
	\bmqty{W_q \delta T_qW_q^{-1} & W_q \Xc_q W^\d_q \\ 
	O & W^{-1,\d}_q \delta T_q^\d W^\d_q
	}_{t_1}^{ij} \;\delta(t_1 - t_2)\,.
\end{align}

The loop contributions of the self-energies can be calculated according to a Feynman diagram
approach. The unperturbed propagators are $4 \times 4$-matrices, and thus the calculations are
complicated and lengthy. The use of a tensor form, which was 
developed for the $2\times 2$ TFD formulation in Ref.~\cite{ChuTensor} 
and was also used for condensate systems in Ref.~\cite{AnnPhys325}, is extremely helpful
in calculating the loop contributions of 
the full propagators and self-energies in a systematic and concise manner.
Although the present calculations need a four-component tensor,
the rules of the tensor calculus are essentially independent of the number of tensor
components, the details of which are found in Refs.~\cite{ChuTensor,AnnPhys325}. 
Because the explicit forms of the calculated loop self-energies 
are not important in our discussions below, herein we give only 
examples of $\Sigma_{\mathrm {int},\bq}^{ij}$, and $S_{\mathrm {int},\bq}^{ij}$, i.e., 
one-loop contributions, which originate from the cubic terms 
in $\HH_{\mathrm {int}}$\,, 
\begin{align}
&\Sigma^{ij}_\bq \DiagramA(t_1,t_2)
=\frac{2ig^2 n_c}{V}\sum_{i_1,i_2}\sum_{\bq_1,\bq_2}
\delta_{\bq,\bq_1+\bq_2} \notag\\
&\quad \times\left\{(\Cc_{03}\Bc_{q_1}^{-1})^{i i_1}\Bc_{q_2}^{-1,i i_2}
+(\Cc_{03}\Bc_{q_1}^{-1})^{i i_1} (\Cc_{11}\Bc_{q_2}^{-1})^{i i_2}
+(\Cc_{11}\Bc_{q_1}^{-1})^{ii_1}(\Cc_{03}\Bc_{q_2}^{-1})^{i i_2}
\right\}_{t_1} \notag \\
&\quad  \times  
d_{\bq_1}^{i_1i_1}(t_1,t_2)d_{\bq_2}^{i_2i_2}(t_1,t_2) \notag \\
&\quad \times \left\{
(\Bc_{q_1}\Cc_{33})^{i_1 j}\Bc_{q_2}^{i_2 j}
-(\Bc_{q_1}\Cc_{33})^{i_1 j}(\Bc_{q_2}\Cc_{11})^{i_2j}
-(\Bc_{q_1}\Cc_{11})^{i_1 j}(\Bc_{q_2}\Cc_{33})^{i_2 j}
\right\}_{t_2}\,. \label{eq:oneloopsigma}
\end{align}
The symbol $\Cc_{\ell_1 \ell_2}$ stands for $(\tau_{\ell_1}\otimes \sigma_{\ell_2})$,
where $\tau_\ell$ and $\sigma_\ell$ $(\ell=1,2,3)$ are the Pauli matrices,
and $\tau_0$ and $\sigma_0$ are the unit matrices, which are explicitly written as
\begin{align}
\Cc_{11}= \bmqty{O&\sigma_1 \\ \sigma_1&O }\,,
\quad \Cc_{03}= \bmqty{\sigma_3& O\\ O& \sigma_3 }\,,
\quad \Cc_{33}= \bmqty{\sigma_3& O\\ O&- \sigma_3 }\,.
\end{align}
From Eq.~(\ref{xi-selfenergy}),
we obtain
\begin{align}
&S_{\bq}^{ij}\DiagramA(t_1,t_2)=\frac{2ig^2 n_c}{V}\sum_{i_1,i_2, i',j'}\sum_{\bq_1,\bq_2}
\delta_{\bq,\bq_1+\bq_2}\, \Bc^{ii'}_q(t_1)\notag\\
&\quad \times \left\{(\Cc_{03}\Bc_{q_1}^{-1})^{i' i_1}\Bc_{q_2}^{-1,i' i_2}
+(\Cc_{03}\Bc_{q_1}^{-1})^{i' i_1} (\Cc_{11}\Bc_{q_2}^{-1})^{i' i_2}
+(\Cc_{11}\Bc_{q_1}^{-1})^{i'i_1}(\Cc_{03}\Bc_{q_2}^{-1})^{i' i_2}
\right\}_{t_1} \notag \\
&\quad  \times  
d_{\bq_1}^{i_1i_1}(t_1,t_2)d_{\bq_2}^{i_2i_2}(t_1,t_2) \notag \\
&\quad \times \left\{
(\Bc_{q_1}\Cc_{33})^{i_1 j'}\Bc_{q_2}^{i_2 j'}
-(\Bc_{q_1}\Cc_{33})^{i_1 j'}(\Bc_{q_2}\Cc_{11})^{i_2j'}
-(\Bc_{q_1}\Cc_{11})^{i_1 j'}(\Bc_{q_2}\Cc_{33})^{i_2 j'}
\right\}_{t_2}\notag \\
& \quad \times \Bc^{-1,j' j}(t_2) \,.
\end{align}

In summary, the self-energy $S^{ij}_{\bq}(t_1,t_2)$ is a sum of the loop and 
counter term contributions,
\begin{align}\label{eq:TotalS}
S^{ij}_{\bq}(t_1,t_2)=S_{{\rm int}, \bq}^{ij}(t_1,t_2)+ 
S^{ij}_{Q, \bq}(t_1,t_2)+ S^{ij}_{\delta H, \bq}(t_1,t_2) \,,
\end{align}
and its matrix elements satisfy the following relations. The tilde transformation 
gives $\{a_{H,\bq}^i(t)\}\Tilde{\phantom{i}}= \Cc_{01}^{ii'}a_{H,\bq}^{i'}(t)$, 
$\{\ab_{H,\bq}^j(t)\}\Tilde{\phantom{i}}= \ab_{H,\bq}^{j'}(t)\Cc_{01}^{j'j}$\,,
and 
\begin{align}
	S^{*,ij}_{\bq}(t_1,t_2)=-\Cc_{01}^{ii'}S^{i'j'}_{-\bq}(t_1,t_2)\Cc_{01}^{j'j}\,.
	\label{eq:S-TildeSym}
	\end{align}
Furthermore, the quartets $a_H^i$ and $\ab_H^i$ are related to each other,
as can be seen from their definitions, $a_{H,\bq}^i(t)= -i\Cc_{21}^{ii'} \ab_{H,-\bq}^{i'}(t)$ 
and $\ab_{H,\bq}^i(t)
=-i  a_{H,-\bq}^{j'}(t)\Cc_{21}^{j'j}$\,, 
which derive the following relation.
	\begin{align}
	S^{ij}_{\bq}(t_1,t_2)=-\Cc_{21}^{ii'}S^{ j'i'}_{-\bq}(t_2,t_1)\Cc_{21}^{j'j}\,.
	\label{eq:S-aSym}
	\end{align}

\section{Renormalization condition} \label{Sec:renormalization}

We impose the renormalization condition onto the on-shell self-energy. 
In Refs.~\cite{AnnPhys331,IJMPKuwahara}, we proposed an on-shell self-energy for time-dependent energy eigenvalues in nonequilibrium TFD. 
However, we renormalized only a part of the matrix elements of the on-shell self-energy.
Extending the previous prescriptions, we propose renormalization conditions on all
elements of the on-shell self-energy in the present $4\times4$ formulation.

In zero-temperature and finite-temperature quantum field theory with a time
 translational symmetry, the self-energies and propagators are 
two-variables functions that are not dependent on the two times, $t_1$ and $t_2$, individually, 
and are dependent only on the relative time $\tau=t_1-t_2$\,. Subsequently, the on-shell is naturally 
introduced through the Fourier transformation of 
$\tau$ as 
\begin{align} \label{onshell1}
	\Sb = \int_{-\infty}^{\infty} \dd\tau\; S(\tau) \;e^{i\omega \tau} \,,
\end{align}
with the renormalized energy $\omega$.
Note that the factor $e^{i\omega \tau}$ is considered to originate from
 the time-dependence of the operator that annihilates the vacuum.
In defining the on-shell state for systems in nonequilibrium TFD without a time translational symmetry
and with time-dependent renormalized energy, we required a 
thermal causality in which the macroscopic time-dependent quantities 
such as $n_q(t)$ and $m_q(t)$ affect the microscopic motions only in the future 
but not in the past \cite{AnnPhys331,IJMPKuwahara}. To comfort the thermal 
causality, the time integration in the 
self-energy has been divided into the retarded and advanced parts as 
\begin{align}\label{onshell2}
	\Sb[z; t] = 
	\int_{0}^\infty \dd\tau\;  S(t,t-\tau) \, \expint{i}{t}{t-\tau}{z(s)}{s} +
	\int_{-\infty}^0 \dd\tau\; S(t+\tau,t) \, \expint{i}{t+\tau}{t}{z(s)}{s} \,,
\end{align}
which is a functional depending on a function $z(t)$. 
In Refs.~\cite{AnnPhys331, IJMPKuwahara}, the on-shell self-energy 
was defined by setting $z(t)=\omega(t)$ as
 $\Sb[\omega; t]$, which reduces to 
Eq.~\eref{onshell1} within the equilibrium limit. 

In the present formulation, because the $\xi$-operator is time-independent owing to TDCS,
as in Eq.~\eref{eq:TDSC-dtxi}, we propose an on-shell self-energy 
defined by the obvious choice of $z(t) = 0$,
\begin{align}\label{onshell3}
	\Sb^{ij}_q(t) = 
	\int_{0}^\infty \dd\tau\;  S_q^{ij}(t,t-\tau)  +
	\int_{-\infty}^0 \dd\tau\; S_q^{ij}(t+\tau,t)  \,.
\end{align}
Here and below, we write the momentum index $\bq$ simply as $q$ because it is a function of
$q$.
Notations of the functionals for retarded and 
advanced parts, respectively, are introduced, as in Ref.~\cite{AnnPhys331},
\begin{align} \label{z-representation}
	\Sb^{+, ij}_q(t) = \int_{0}^\infty \dd\tau\;  S^{ij}_q(t,t-\tau) \,,\qquad
	\Sb^{-, ij}_q(t) = \int_{-\infty}^0 \dd\tau\; S^{ij}_q(t+\tau,t)\,.
\end{align}
This and Eqs.~\eref{eq:S-TildeSym} and \eref{eq:S-aSym} yield
\begin{alignat}{7}
	\Sb_q^{+,11}(t) &= 
	&-\bigl(&\Sb_q^{+,22}(t)\bigr)^* &&=
	& \bigl(&\Sb_q^{-,33}(t)\bigr)^* &&=
	&  &-\Sb_q^{-,44}(t)         \,,\label{Ssym0}\\
	\Sb_q^{+,12}(t) &= 
	&-\bigl(&\Sb_q^{+,21}(t)\bigr)^* &&=
	&       &-\Sb_q^{-,34}(t)         &&=
	&\bigl(&\Sb_q^{-,43}(t)\bigr)^* \,,\\
	\Sb_q^{+,13}(t) &= 
	&-\bigl(&\Sb_q^{+,24}(t)\bigr)^* &&=
	&-\bigl(&\Sb_q^{-,13}(t)\bigr)^* &&=
	&       &\Sb_q^{-,24}(t)         \,,\\
	\Sb_q^{+,14}(t) &= 
	&-\bigl(&\Sb_q^{+,23}(t)\bigr)^* &&=
	&       &\Sb_q^{-,14}(t)         &&=
	&-\bigl(&\Sb_q^{-,23}(t)\bigr)^* \,,\label{Ssym3}
\end{alignat}
and we rewrite Eq.~\eref{onshell3} as
\begin{align}
	\Sb^{ij}_q(t) &= 
	\pmqty{\begin{array}{cc|cc}
		\Sb_q^{+,11}(t) &
		\Sb_q^{+,12}(t) &
		2i\Im \Sb_q^{+,13}(t) &
		2\Sb_q^{+,14}(t)  \\[2mm]
		-\qty(\Sb_q^{+,12}(t))^*  &
		-\qty(\Sb_q^{+,11}(t))^* & 
		-2\qty(\Sb_q^{+,14}(t))^*  &
		2i\Im \Sb_q^{+,13}(t)    \\[2mm]\hline
		&&
		\qty(\Sb_q^{+,11}(t))^* &
		-\Sb_q^{+,12}(t) \\[2mm]
		&&
		\qty(\Sb_q^{+,12}(t))^* &
		-\Sb_q^{+,11}(t)
	\end{array}}^{ij}\,.
\end{align}
It can be clearly seen that the on-shell self-energy consists of seven independent real functions:
$\Sb_q^{+,11}(t)$, $\Sb_q^{+,12}(t)$, $\Sb_q^{+,14}(t)$, and $\Im \Sb_q^{+,13}(t)$\,.

We now propose the renormalization condition as 
\begin{align} \label{renoramlizationcondition}
	\Sb_q^{ij}(t) = 0 \,,
\end{align}
which is equivalent to the seven equations, 
\begin{align}
	\Im \Sb_q^{+,13}(t) = \Sb_q^{+,11}(t) = \Sb_q^{+,12}(t) = \Sb_q^{+,14}(t) = 0\,.
\end{align}
Recall that the counter terms have only seven real parameters, i.e., 
three real functions $\delta\Lc_q\,,\, \kappa_q\,,\,\nd_q$ and two complex functions 
$\delta \Mc_q\,,\,\md_q$. We emphasize that the renormalization condition in 
Eq.~\eref{renoramlizationcondition} is a necessary and sufficient 
condition to determine these seven parameters. 

The on-shell of Eq.~\eref{eq:TotalS} is
\begin{align}
	\Sb_q(t) =\Sb_{\mathrm{int},q}(t)+ \Sb_{Q,q}(t)+\Sb_{\delta H,q}(t)  \,,
\end{align} 
where 
\begin{align}
	&\bar{S}_{Q,q}^{ij}(t)= -i\bmqty{  O& W_q\dot{N}_q W_q^\d \\ O & O}_{t}^{ij}\,,\\
	&\bar{S}_{\delta H, q}(t) = \bmqty{W_q \delta T_qW_q^{-1} & W_q \Xc_q W^\d_q \\
		O & W_q^{-1,\d}\delta T_q^\d W_q^\d}_{t}^{ij}
	 \,,
\end{align}
with the definitions of Eqs.~\eref{eq:WN} and \eref{eq:defdeltaL}--\eref{eq:defXc}.
The renormalization condition in Eq.~\eref{renoramlizationcondition} implies
\begin{align}\label{11block}
	\delta T(t)=- W^{-1}_q(t) \pmqty{\Sbi{11}(t) & \Sbi{12}(t)\\ 
				 - \qty(\Sbi{12}(t))^* & - \qty(\Sbi{11}(t))^*}  W_q(t) \,,
\end{align}
and
\begin{align}	
	\dot{N}_q(t) = - 
	2iW^{-1}_q(t) \pmqty{ i\Im \Sbi{13}(t) & \Sbi{14}(t) \\ 
		-\qty(\Sbi{14}(t))^* & i \Im \Sbi{13}(t) } W^{\d,-1}_q(t)- i\Xc_q(t) \,.
\label{12block}
\end{align}
Decomposing the matrices, we finally obtain
\begin{align}
	&\delta \Lc_q =  2\Re \qty[ c_{q} s_{q} \Sbi{12}] - \qty(|c_{q}|^2 + |s_{q}|^2)\Re 
\Sbi{11} \,,\label{eq:RendeltaL}\\
	&\kappa_q 	= \Im \Sbi{11} \label{eq:Renkappa}\,,\\
	&\delta\Mc_q	= 2c_{q} s_{q}^* \Re \Sbi{11} -c_{q}^2 \Sbi{12} + \qty(s^{2}_{q} \Sbi{12})^* \,,\label{eq:RendeltaM}\\
	&\nd_q = 2\qty(|c_q|^2 + |s_q|^2)\Im \Sbi{13} + 
	4\Im\qty[c_q s_q \Sbi{14}] + 2\Im \qty[m_q gn_c] \,,\\
	&\md_q = 4c_qs^*_q \Im \Sbi{13} -2ic^2_q \Sbi{14} + 2i \qty(s^{2}_q \Sbi{14})^* 
	+ i(1+2n_q)gn_c+ 2i m_q (\varepsilon_q+gn_c)\,.\label{eq:Renmdot}
\end{align}
In this way, the last two equations, i.e., 
the quantum transport equations for normal and anomalous density
distributions, are derived. Equations~\eref{eq:RendeltaL} 
and \eref{eq:RendeltaM} represent the energy shift.
As an important result, $\kappa_q$, which represents the thermal instability 
whose inverse is a measure of the relaxation time to equilibrium, is also
determined through the renormalization.

\section{Conclusion and summary}\label{sec-Conclusion}

In this paper, we have constructed a consistent TFD formulation for a nonequilibrium 
quantum field system of a homogeneous BEC associated with an SSB. To handle such thermal situations, 
in the unperturbed hat Hamiltonian $\HH_a$\, we need phase symmetry-breaking terms 
such as $aa$, in addition to the cross-terms between non-tilde and tilde operators such as $a\at$\,.
Thus, the general thermal Bogoliubov transformation used to relate $a$-operators with 
$\xi$-operators that
define the thermal vacua and diagonalize $\HH_a$ is represented using a $4\times4$-matrix.
We require the minimal conditions on a bilinear $\HH_a$, 
through Eqs.~\eref{HaCond1} and \eref{HaCond2}, related to the tilde conjugation and the $\alpha = 1$ representation, respectively.
It was shown that there are seven independent real parameters in $\HH_a$. 
The self-energy $4\times 4$-matrix of 
the loop and counter term contributions can be calculated according to the Feynman diagram method. 
A crucial step in this study is the use of the time-dependent BdG equation,  
which makes the thermal quasiparticle operators time-independent
such that the on-shell part of the self-energy can be defined without ambiguity. The renormalization 
conditions in Eq.~\eref{renoramlizationcondition}, implying that the entire  
on-shell self-energy should vanish, yield seven independent equations
to determine 
the seven parameters in $\HH_a$\, without excess or deficiency, as in
 Eqs. \eref{eq:RendeltaL}--\eref{eq:Renmdot}. Two of these are the quantum transport equations. 
The others fix the five parameters of the time-dependent BdG equations, corresponding to
 the energy shift and the thermal instability.

We have treated similar nonequilibrium systems of BEC within TFD \cite{AnnPhys325, AnnPhys326}.
We then assumed sequential $2\times2$ transformations, usual and thermal, 
instead of a general $4\times4$ transformation. It can be shown that 
the $4\times 4$ transformation 
obtained in this study can also be put into a form of two sequential $2\times2$ 
transformations. On the other hand, 
the unperturbed hat Hamiltonians in Refs.~\cite{AnnPhys325, AnnPhys326} differ 
 from the unperturbed hat Hamiltonian in this study in such a manner 
that the parameter of the thermal instability 
$\kappa$ vanishes for the former, but does not vanish for the latter.
The difference is as follows: In previous studies, an 
intermediate quasi-particle operator, 
say a $b$-operator, was introduced and two-step transformations,
\begin{align}
\xi = \Bc_{\mathrm{thermal}} b \,,\qquad 
b = \Bc_{\mathrm{usual}} a\,,\quad
\Bc_{\mathrm{thermal}} \Bc_{\mathrm{usual}} = \Bc \,.
\end{align}
were performed. Finally, the unperturbed hat Hamiltonian was required
to be diagonal in $b$-operators, which constrained $\kappa = 0$\,. 
The presence of $\kappa$, which makes the unprepared Hamiltonian non-Hermitian 
[see Eq.~\eref{diagonalizedHx}], may become a major obstacle to conventional 
quantum field theory, whereas nonequilibrium TFD can take it in without a contradiction.

The derivation of the quantum transport equations in this paper remains formal. 
We  presented only one-loop self-energy in Eq.~\eref{eq:oneloopsigma}.
However, the equations have not yet been analyzed numerically. 
In addition, the role of $\kappa$ in the transport equations should be clarified.  
These remain as future tasks.

The study with several restrictions in this paper 
is the first step of a general 
$4\times4$ formulation. The restrictions can be lifted in the following way.
 We assumed a system with a homogeneous and stationary condensate,
for which the momentum is a good quantum number and the unperturbed hat Hamiltonian 
is given as a simple sum of that of a single mode. In principle, it is  not 
difficult to
extend it to a system with an inhomogeneous and non-stationary condensate. 
The basic ideas are presented in Refs.~\cite{AnnPhys325, AnnPhys326, IJMPKuwahara}. 
In addition, we considered only non-zero real $\omega$, defined 
just below Eq.~\eref{eq:defEyy} and 
 after Eq.~\eref{lrBdG}. 
In the case of a complex $\omega$, implying dynamical instability of the condensate, 
the bilinear Hamiltonian cannot be diagonalized by any particle operator with 
a bosonic commutation relation even at zero temperature \cite{Mine2007}. 
In case of a vanishing $\omega$, $T$ is also non-diagonalizable 
\cite{LewensteinYou, MatsumotoSakamoto, Mine2005}. An important example is 
the Nambu--Goldstone mode \cite{Nambu, Goldstone} in the SSB, which has typically been neglected 
as a Bogoliubov approximation \cite{PethickSmith} for homogeneous systems. 
We considered inhomogeneous, finite systems, for which the zero mode appears as a discrete level and the Bogoliubov approximation is inapplicable, and proposed a formulation 
beyond the Bogoliubov approximation at zero temperature \cite{NTY}. 
The development of the $4\times4$-matrix formulation of nonequilibrium TFD 
for a dynamically unstable system and for inhomogeneous system with a discrete Nambu--Goldstone
mode will be a challenge.  

\section*{Acknowledgements}
The authors thank KEK Theory Center for offering us the opportunity to 
discuss this work during the KEK Theory workshop 
 on ``Thermal Quantum Field Theories and Their Applications,'' and
Dr.~R.~Imai for the discussions conducted at an early stage of this study.



\section*{References}
\bibliographystyle{model1a-num-names}
\bibliography{ref44TFD}

\begin{thebibliography}{10}
\expandafter\ifx\csname url\endcsname\relax
  \def\url#1{\texttt{#1}}\fi
\expandafter\ifx\csname urlprefix\endcsname\relax\def\urlprefix{URL }\fi
\expandafter\ifx\csname href\endcsname\relax
  \def\href#1#2{#2} \def\path#1{#1}\fi

\bibitem{PethickSmith}
C.~J. Pethick, H.~Smith, Bose--Einstein condensation in dilute gases, Cambridge
  university press, 2008.
\newblock \href {http://dx.doi.org/10.1017/CBO9780511802850}
  {\path{doi:10.1017/CBO9780511802850}}.

\bibitem{GriffinBook}
A.~Griffin, T.~Nikuni, E.~Zaremba, Bose-condensed gases at finite temperatures,
  Cambridge University Press, 2009.
\newblock \href {http://dx.doi.org/10.1017/CBO9780511575150}
  {\path{doi:10.1017/CBO9780511575150}}.

\bibitem{VitielloBook}
M.~Blasone, G.~Vitiello, P.~Jizba, Quantum Field Theory and its macroscopic
  manifestations: Boson Condensation, Ordered Patterns and Topological Defects,
  Imperial College Press, 2011.

\bibitem{Jin}
D.~Jin, M.~Matthews, J.~Ensher, C.~Wieman, E.~A. Cornell, Temperature-dependent
  damping and frequency shifts in collective excitations of a dilute
  bose-einstein condensate, Physical Review Letters 78~(5) (1997) 764.
\newblock \href {http://dx.doi.org/10.1103/PhysRevLett.78.764}
  {\path{doi:10.1103/PhysRevLett.78.764}}.

\bibitem{Yamashita}
M.~Yamashita, M.~Koashi, N.~Imoto, Quantum kinetic theory for evaporative
  cooling of trapped atoms: Growth of bose-einstein condensate, Physical Review
  A 59~(3) (1999) 2243.
\newblock \href {http://dx.doi.org/10.1103/PhysRevA.59.2243}
  {\path{doi:10.1103/PhysRevA.59.2243}}.

\bibitem{Morgan1}
S.~Morgan, M.~Rusch, D.~Hutchinson, K.~Burnett, Quantitative test of thermal
  field theory for bose-einstein condensates, Physical review letters 91~(25)
  (2003) 250403.
\newblock \href {http://dx.doi.org/10.1103/PhysRevLett.91.250403}
  {\path{doi:10.1103/PhysRevLett.91.250403}}.

\bibitem{Morgan2}
S.~Morgan, Quantitative test of thermal field theory for bose-einstein
  condensates. ii, Physical Review A 72~(4) (2005) 043609.
\newblock \href {http://dx.doi.org/10.1103/PhysRevA.72.043609}
  {\path{doi:10.1103/PhysRevA.72.043609}}.

\bibitem{Bezett}
A.~Bezett, P.~Blakie, Projected gross-pitaevskii equation theory of
  finite-temperature collective modes for a trapped bose gas, Physical Review A
  79~(2) (2009) 023602.
\newblock \href {http://dx.doi.org/10.1103/PhysRevA.79.023602}
  {\path{doi:10.1103/PhysRevA.79.023602}}.

\bibitem{Davis}
K.~B. Davis, M.-O. Mewes, M.~R. Andrews, N.~J. van Druten, D.~S. Durfee,
  D.~Kurn, W.~Ketterle, Bose-einstein condensation in a gas of sodium atoms,
  Physical review letters 75~(22) (1995) 3969.
\newblock \href {http://dx.doi.org/10.1103/PhysRevLett.75.3969}
  {\path{doi:10.1103/PhysRevLett.75.3969}}.

\bibitem{Miesner}
H.-J. Miesner, D.~Stamper-Kurn, M.~Andrews, D.~Durfee, S.~Inouye, W.~Ketterle,
  Bosonic stimulation in the formation of a bose-einstein condensate, Science
  279~(5353) (1998) 1005--1007.
\newblock \href {http://dx.doi.org/10.1126/science.279.5353.1005}
  {\path{doi:10.1126/science.279.5353.1005}}.

\bibitem{Sommer}
A.~Sommer, M.~Ku, G.~Roati, M.~W. Zwierlein, Universal spin transport in a
  strongly interacting fermi gas, Nature 472~(7342) (2011) 201--204.
\newblock \href {http://dx.doi.org/10.1038/nature09989}
  {\path{doi:10.1038/nature09989}}.

\bibitem{LesHouches}
T.~Giamarchi, A.~J. Millis, O.~Parcollet, H.~Saleur, L.~F. Cugliandolo,
  Strongly Interacting Quantum Systems out of Equilibrium: Lecture Notes of the
  Les Houches Summer School: Volume 99, August 2012, Vol.~99, Oxford University
  Press, 2016.
\newblock \href {http://dx.doi.org/10.1093/acprof:oso/9780198768166.001.0001}
  {\path{doi:10.1093/acprof:oso/9780198768166.001.0001}}.

\bibitem{Schwinger}
J.~Schwinger, Brownian motion of a quantum oscillator, Journal of Mathematical
  Physics 2~(3) (1961) 407--432.
\newblock \href {http://dx.doi.org/10.1063/1.1703727}
  {\path{doi:10.1063/1.1703727}}.

\bibitem{Keldysh}
L.~V. Keldysh, et~al., Diagram technique for nonequilibrium processes, Sov.
  Phys. JETP 20~(4)  1018--1026.

\bibitem{KadanoffBaym}
L.~Kadanoff, G.~Baym, Quantum statistical mechanics, CRC Press, 1962.

\bibitem{Danielewiecz}
P.~Danielewicz, Quantum theory of nonequilibrium processes, i, Annals of
  Physics 152~(2) (1984) 239--304.
\newblock \href {http://dx.doi.org/10.1016/0003-4916(84)90092-7}
  {\path{doi:10.1016/0003-4916(84)90092-7}}.

\bibitem{Chou}
K.-c. Chou, Z.-b. Su, B.-l. Hao, L.~Yu, Equilibrium and nonequilibrium
  formalisms made unified, Physics Reports 118~(1-2) (1985) 1--131.
\newblock \href {http://dx.doi.org/10.1016/0370-1573(85)90136-X}
  {\path{doi:10.1016/0370-1573(85)90136-X}}.

\bibitem{UMT}
H.~Umezawa, H.~Matsumoto, M.~Tachiki, Thermo Field Dynamics and Condensed
  States, North-Holland, 1982.

\bibitem{AIP}
H.~Umezawa, Advanced Field Theory --- Micro, Macro, and Thermal Physics, AIP,
  1993.

\bibitem{AnnPhys325}
Y.~Nakamura, T.~Sunaga, M.~Mine, M.~Okumura, Y.~Yamanaka, Derivation of
  non-markovian transport equations for trapped cold atoms in nonequilibrium
  thermal field theory, Annals of Physics 325~(2) (2010) 426--441.
\newblock \href {http://dx.doi.org/10.1016/j.aop.2009.09.014}
  {\path{doi:10.1016/j.aop.2009.09.014}}.

\bibitem{ChuUmezawa}
H.~Chu, H.~Umezawa, Renormalization and boltzmann equations in thermal quantum
  field theory, International Journal of Modern Physics A 10~(11) (1995)
  1693--1700.
\newblock \href {http://dx.doi.org/10.1142/S0217751X95000814}
  {\path{doi:10.1142/S0217751X95000814}}.

\bibitem{AnnPhys326}
Y.~Nakamura, Y.~Yamanaka, Unifying treatment of nonequilibrium and unstable
  dynamics of cold bosonic atom system with time-dependent order parameter in
  thermo field dynamics, Annals of Physics 326~(4) (2011) 1070--1083.
\newblock \href {http://dx.doi.org/10.1016/j.aop.2010.12.002}
  {\path{doi:10.1016/j.aop.2010.12.002}}.

\bibitem{AnnPhys331}
Y.~Nakamura, Y.~Yamanaka, From superoperator formalism to nonequilibrium thermo
  field dynamics, Annals of Physics 331 (2013) 51--69.
\newblock \href {http://dx.doi.org/10.1016/j.aop.2012.12.005}
  {\path{doi:10.1016/j.aop.2012.12.005}}.

\bibitem{IJMPKuwahara}
Y.~Kuwahara, Y.~Nakamura, Y.~Yamanaka, Self-energy renormalization for
  inhomogeneous nonequilibrium systems and field expansion via complete set of
  time-dependent wavefunctions, International Journal of Modern Physics B
  32~(10) (2018) 1850111.
\newblock \href {http://dx.doi.org/10.1142/S0217979218501114}
  {\path{doi:10.1142/S0217979218501114}}.

\bibitem{Elmfors}
P.~Elmfors, H.~Umezawa, Generalizations of the thermal bogoliubov
  transformation, Physica A: Statistical Mechanics and its Applications
  202~(3-4) (1994) 577--594.
\newblock \href {http://dx.doi.org/10.1016/0378-4371(94)90480-4}
  {\path{doi:10.1016/0378-4371(94)90480-4}}.

\bibitem{Matsumoto2001}
H.~Matsumoto, S.~Sakamoto, Nonequilibrium formulation in bose-einstein
  condensed states, Progress of Theoretical Physics 105~(4) (2001) 573--590.
\newblock \href {http://dx.doi.org/10.1143/PTP.105.573}
  {\path{doi:10.1143/PTP.105.573}}.

\bibitem{NTY}
Y.~Nakamura, J.~Takahashi, Y.~Yamanaka, Formulation for the zero mode of a
  bose-einstein condensate beyond the bogoliubov approximation, Physical Review
  A 89~(1) (2014) 013613.
\newblock \href {http://dx.doi.org/10.1103/PhysRevA.89.013613}
  {\path{doi:10.1103/PhysRevA.89.013613}}.

\bibitem{SchmutzSO}
M.~Schmutz, Real-time green's functions in many body problems, Zeitschrift
  f{\"u}r Physik B Condensed Matter 30~(1) (1978) 97--106.
\newblock \href {http://dx.doi.org/10.1007/BF01323673}
  {\path{doi:10.1007/BF01323673}}.

\bibitem{Evans}
T.~Evans, I.~Hardman, H.~Umezawa, Y.~Yamanaka, Heisenberg and interaction
  representations in thermo field dynamics, Journal of mathematical physics
  33~(1) (1992) 370--378.
\newblock \href {http://dx.doi.org/10.1063/1.529915}
  {\path{doi:10.1063/1.529915}}.

\bibitem{Bogoliubov}
N.~Bogoliubov, On the theory of superfluidity, J. Phys 11~(1)  23.

\bibitem{deGennes}
P.-G. De~Gennes, Superconductivity of metals and alloys, Benjamin, 1966.

\bibitem{Fetter}
A.~L. Fetter, Nonuniform states of an imperfect bose gas, Annals of Physics
  70~(1) (1972) 67--101.
\newblock \href {http://dx.doi.org/10.1016/0003-4916(72)90330-2}
  {\path{doi:10.1016/0003-4916(72)90330-2}}.

\bibitem{Chen}
T.-W. Chen, S.-C. Cheng, W.-F. Hsieh, Collective excitations, nambu-goldstone
  modes, and instability of inhomogeneous polariton condensates, Physical
  Review B 88~(18) (2013) 184502.
\newblock \href {http://dx.doi.org/10.1103/PhysRevB.88.184502}
  {\path{doi:10.1103/PhysRevB.88.184502}}.

\bibitem{ChuTensor}
H.~Chu, H.~Umezawa, Time ordering theorem and calculational recipes for thermo
  field dynamics, Physics Letters A 177~(6) (1993) 385--393.
\newblock \href {http://dx.doi.org/10.1016/0375-9601(93)90963-Z}
  {\path{doi:10.1016/0375-9601(93)90963-Z}}.

\bibitem{Mine2007}
M.~Mine, M.~Okumura, T.~Sunaga, Y.~Yamanaka, Quantum field theoretical
  description of unstable behavior of trapped bose--einstein condensates with
  complex eigenvalues of bogoliubov--de gennes equations, Annals of Physics
  322~(10) (2007) 2327--2349.
\newblock \href {http://dx.doi.org/10.1016/j.aop.2007.01.008}
  {\path{doi:10.1016/j.aop.2007.01.008}}.

\bibitem{LewensteinYou}
M.~Lewenstein, L.~You, Quantum phase diffusion of a bose-einstein condensate,
  Physical Review Letters 77~(17) (1996) 3489.
\newblock \href {http://dx.doi.org/10.1103/PhysRevLett.77.3489}
  {\path{doi:10.1103/PhysRevLett.77.3489}}.

\bibitem{MatsumotoSakamoto}
H.~Matsumoto, S.~Sakamoto, Quantum phase coordinate as a zero-mode in
  bose-einstein condensed states, Progress of theoretical physics 107~(4)
  (2002) 679--688.
\newblock \href {http://dx.doi.org/10.1143/PTP.107.679}
  {\path{doi:10.1143/PTP.107.679}}.

\bibitem{Mine2005}
M.~Mine, M.~Okumura, Y.~Yamanaka, Relation between generalized bogoliubov and
  bogoliubov--de gennes approaches including nambu--goldstone mode, Journal of
  mathematical physics 46~(4) (2005) 042307.
\newblock \href {http://dx.doi.org/10.1063/1.1865322}
  {\path{doi:10.1063/1.1865322}}.

\bibitem{Nambu}
Y.~Nambu, G.~Jona-Lasinio, Dynamical model of elementary particles based on an
  analogy with superconductivity. i, Physical review 122~(1) (1961) 345.
\newblock \href {http://dx.doi.org/10.1103/PhysRev.122.345}
  {\path{doi:10.1103/PhysRev.122.345}}.

\bibitem{Goldstone}
J.~Goldstone, Field theories with superconductor solutions, Il Nuovo Cimento
  (1955-1965) 19~(1) (1961) 154--164.

\end{thebibliography}

\end{document}